\documentclass[twocolumn,superscriptaddress,showpacs,preprintnumbers,amsmath,amssymb]{revtex4-1}

\usepackage{amsmath}
\usepackage{amssymb}
\usepackage{graphicx}
\usepackage{morefloats}
\usepackage[usenames,dvipsnames]{xcolor}
\usepackage{blkarray}
\usepackage{verbatim}
\usepackage{hyperref}
\usepackage{mathtools}

\numberwithin{equation}{section}
\hypersetup{colorlinks=true, citecolor=orange, urlcolor=cyan, linkcolor=blue}

\begin{document}

\title{Reconstructing mesoscale network structures}
\author{Jeroen van Lidth de Jeude}
\affiliation{IMT School for Advanced Studies, Piazza S.Francesco 19, 55100 Lucca - Italy}
\author{Riccardo Di Clemente}
\affiliation{University College London, The Bartlett Centre for Advanced Spatial Analysis, Gower Street, WC1E 6BT, London - United Kingdom}
\author{Guido Caldarelli}
\affiliation{IMT School for Advanced Studies, Piazza S.Francesco 19, 55100 Lucca - Italy}
\author{Fabio Saracco}
\affiliation{IMT School for Advanced Studies, Piazza S.Francesco 19, 55100 Lucca - Italy}
\author{Tiziano Squartini}
\affiliation{IMT School for Advanced Studies, Piazza S.Francesco 19, 55100 Lucca - Italy}
\date{\today}

\begin{abstract}
When facing the problem of reconstructing complex mesoscale network structures, it is generally believed that models encoding the nodes organization into modules must be employed. The present paper focuses on two block structures that characterize the empirical mesoscale organization of many real-world networks, i.e. the \emph{bow-tie} and the \emph{core-periphery} ones, with the aim of quantifying the minimal amount of topological information that needs to be enforced in order to reproduce the topological details of the former. Our analysis shows that constraining the network degree sequences is often enough to reproduce such structures, as confirmed by model selection criteria as AIC or BIC. As a byproduct, our paper enriches the toolbox for the analysis of bipartite networks - still far from being complete: both the bow-tie and the core-periphery structure, in fact, partition the networks into asymmetric blocks characterized by binary, directed connections, thus calling for the extension of a recently-proposed method to randomize {\it undirected}, bipartite networks to the \emph{directed} case.
\end{abstract}

\maketitle

\section{Introduction\label{intro}}

The analysis of mesoscale network structures is a topic of great interest within the community of network scientists: much attention, however, has been received by the community-detection topic \cite{Fortunato2010,Fortunato2016,Bisma2017}, while the analysis of other meso-structures has remained far less explored.

The present work aims at contributing to this stream of research, by exploring the effectiveness of models that constrain only local information in reproducing complex meso-structures as the bow-tie and the core-periphery ones. When approaching such a problem it is, in fact, commonly believed that models encoding the nodes organization into modules must be employed: here we test this hypothesis, by comparing models that enforce topological information like the total number of links, the degree sequences and the reciprocity structure with their block-wise counterparts.

To this aim, we have considered real-world networks whose topological structure is \emph{empirically} characterized by bow-tie and core-periphery structures: both are characterized by a central, cohesive subgraph surrounded by a loosely-connected set of nodes \cite{Csermely2013}; in the first case, however, the central part of the network has a fan-in and a fan-out component, respectively entering into and exiting from it.

Remarkably, all models considered in the present paper can be recovered within the same framework, i.e. the entropy-maximization one, which has been proven to be rather effective for approaching both pattern detection and real-world networks reconstruction problems \cite{Squartini2011,Mastrandrea2014a}. Such a framework allows a tunable likelihood function to be definable for each considered model, thus allowing selection criteria like AIC or BIC to be applicable for unambiguously determining the ``winner'' between competing models, i.e. the one carrying the right amount of information to account for the inspected structures.

As a byproduct, our paper enriches the toolbox for the analysis of bipartite networks. Among the many, available, network representations, the bipartite one has recently received much attention \cite{Tacchella2012,Cimini2014}. This, in turn, has led to the definition of algorithms for randomizing \cite{Kitsak2011,Dormann2014,Strona2014,Saracco2015}, reconstructing \cite{Squartini2017} or projecting \cite{Tummi2011,Saracco2017} {\it undirected}, bipartite networks. Their directed representation, however, has not been explored yet, thus calling for the definition of techniques to approach the study of this kind of networks as well.

This is especially true when considering that bipartite networks emerge quite naturally when studying the aforementioned mesoscale structures. It is, in fact, evident that analysing the way nodes cluster together unavoidably leads to the analysis of the way such modules interact. From an algebraic point of view, this boils down to consider matrices characterized by diagonal square blocks (i.e. the adjacency matrices of the modules themselves) and off-diagonal rectangular blocks (i.e. the adjacency matrices of the bipartite networks encoding their interactions).

Our method will be employed to analyse economic and financial networks empirically characterized by either bow-tie or core-periphery structures: more specifically, we will focus on two systems, the World Trade Web and the Dutch Interbank Network. As we will show, while the former can be described by a partial bow-tie structure, the latter is characterized by the co-existence of a core-periphery\:-like structure and a proper bow-tie one, the second one carrying a larger amount of information about the system evolution than the first one.

\section{Data\label{data}}

Let us now describe the two systems we have considered for the present analysis.\\

\paragraph*{The World Trade Web.} We consider yearly bilateral data on exports and imports from the UN COMTRADE database \cite{uncom}, from 1992 to 2002. We limit ourselves to considering the World Trade Web (WTW hereafter) in its binary, directed representation at the aggregate level. In order to perform a temporal analysis and compare different years, we restrict ourselves to a balanced panel of $N=162$ countries (present in the data throughout the considered interval). Accordingly, for a given year $t$, $a_{ij}^t=1$ ($a_{ij}^t=0$) means that country $i$ has registered a non-null (null) export towards country $j$.\\

\paragraph*{The Dutch Interbank Network.} We consider a dataset where nodes are Dutch banks and a link from node $i$ to node $j$ indicates that bank $i$ has an exposure larger than 1.5 million euros and with maturity shorter than one year, towards a creditor bank $j$ \cite{Lelyveld2012}. We consider 44 quarterly snapshots of the Dutch Interbank Network (DIN hereafter), from 1998Q1 to 2008Q4. The last year in the sample represents the year during which the recent financial crisis became manifest.

\section{Methods\label{methods}}

\subsection{The general framework}

Let us, first, provide an algebraic representation of the mesoscale structures considered in the present paper, i.e. the bow-tie and the core-periphery ones.

Networks whose topology is empirically characterized by a core-periphery structure can be represented as follows:

\begin{equation}\label{cp}
\mathbf{A}=\left( \begin{array}{cc}
\mathbf{A}^\bullet & \mathbf{A}^\top\\
\mathbf{A}^\bot & \mathbf{A}^\circ
\end{array} \right);
\end{equation}
the adjacency matrix $\mathbf{A}$ is composed by four distinct blocks: while the square adjacency matrices $\mathbf{A}^\bullet$ and $\mathbf{A}^\circ$ lying along the diagonal represent the core and the periphery modules, the two rectangular (in the most general case), off-diagonal matrices $\mathbf{A}^\top$ and $\mathbf{A}^\bot$ represent the (bipartite) networks through which they interact. Usually, the link densities of the matrices above satisify the chain of relationships $c(\mathbf{A}^\bullet)>c(\mathbf{A}^\top)\simeq c(\mathbf{A}^\bot)>c(\mathbf{A}^\circ)$, i.e. the core module is (much) denser than the periphery module.

Notice that the two matrices $\mathbf{A}^\top$ and $\mathbf{A}^\bot$ bring genuinely different information: while the generic entry $a_{cp}^\top=1$ ($a_{cp}^\top=0$) indicates that a directed link from the node $c$ in the core to the node $p$ in the periphery is present (absent), the generic entry $a_{pc}^\bot=1$ ($a_{pc}^\bot=0$) indicates that a directed link from the periphery node $p$ to the core node $c$ is present (absent). In other words, in order to fully describe the topological structure of \emph{one}, \emph{directed} bipartite network, \emph{two} matrices are, in fact, needed. Naturally, in case the network $\mathbf{A}$ is undirected, $\mathbf{A}^\bullet=[\mathbf{A}^\bullet]^{\text{T}}$, $\mathbf{A}^\circ=[\mathbf{A}^\circ]^{\text{T}}$ and $\mathbf{A}^\top=[\mathbf{A}^\bot]^{\text{T}}$, which restores the symmetry of the whole adjacency matrix (i.e. $\mathbf{A}=\mathbf{A}^{\text{T}}$).\\

While the definition of core-periphery structure is quite intuitive, the definition of bow-tie structure, on the other hand, is based on the concept of node \emph{reachability}: node $i$ is reachable from node $j$ if a path exists from node $i$ to node $j$ (a path being defined as a sequence of adjacent links connecting $i$ with $j$). According to this definition, each node is assigned to one of the sets described in \cite{Yang2011}. The definition of the three most relevant ones follows:

\begin{itemize}
\item SCC: each node in the Strongly Connected Component (SCC) is reachable from any other node belonging to the SCC;
\item IN: each node in the SCC is reachable from any node belonging to the IN-component;
\item OUT: each node in OUT-component is reachable from any node belonging to the SCC.
\end{itemize}

According to the definitions above, networks whose topology is empirically characterized by a bow-tie structure can be represented by the following adjacency matrix

\begin{equation}\label{bt}
\mathbf{A}=\left( \begin{array}{ccc}
\mathbf{A}^{I} & \mathbf{A}^\rangle & \mathbf{0}\\
\mathbf{0} & \mathbf{A}^{S} & \mathbf{A}^{\rangle\rangle}\\
\mathbf{0} & \mathbf{0} & \mathbf{A}^{O}
\end{array} \right)
\end{equation}
the three blocks $\mathbf{A}^{S}$, $\mathbf{A}^{I}$ and $\mathbf{A}^{O}$ representing the SCC, IN- and OUT-component respectively. The off-diagonal matrices $\mathbf{A}^\rangle$ and $\mathbf{A}^{\rangle\rangle}$, instead, represent the (bipartite) networks through which they interact.

\subsection{Null models}

Let us now provide a brief description of the set of models that will be implemented to analyse the two kinds of mesoscale structures described above (for a detailed description see Appendix A). Let us also clarify that we will proceed by comparing the empirical network structures with models that constrain an increasing amount of information: in other words, we will compare our observations with increasingly refined benchmarks, a way of proceeding that justifies our choice of naming the latter \emph{null models}.\\

The first class of null models we consider for the present analysis is the one including the so-called \emph{degree-informed null models}. All null models in this class are defined by constraints encoding node-specific local information (i.e. the directed degree sequences), beside the membership of nodes to specified groups (labeled by the symbols $\{g_i\}$). Upon combining these two kinds of information, one obtains, in the most general case, block-specific directed degree sequences, definable as 

\begin{eqnarray}
k_i^{r\rightarrow s}&=&\delta_{g_ir}\sum_{j(\neq i)}\delta_{g_js}a_{ij},\:\forall\:i,r,s\\
h_i^{s\rightarrow r}&=&\delta_{g_ir}\sum_{j(\neq i)}\delta_{g_js}a_{ji},\:\forall\:i,r,s
\end{eqnarray}
with $k_i^{r\rightarrow s}$ indicating the contribution to the out-degree of node $i$ (belonging to block $r$) coming from block $s$ (and analogously for $h_i^{s\rightarrow r}$). Remarkably, all null models in this class induce a probability for the generic network configuration $\mathbf{A}$ reading

\begin{equation}
P(\mathbf{A})=\prod_{i\neq j}p_{ij}^{a_{ij}}(1-p_{ij})^{1-a_{ij}}
\end{equation}
with $p_{ij}$ being (in the most general case)

\begin{equation}
p_{ij}=\frac{x_i^{g_i\rightarrow g_j}y_j^{g_i\rightarrow g_j}}{1+x_i^{g_i\rightarrow g_j}y_j^{g_i\rightarrow g_j}},
\label{pij}
\end{equation}
an expression making the dependence of the nodes degree(s) on the group membership apparent. Notice that all degree-informed null models considered here can be recovered from eq. \ref{pij} upon opportunely relaxing the aforementioned dependencies. As an example, the directed version of the Stochastic Block Model (SBM) can be recovered by posing $x_i^{g_i\rightarrow g_j}y_j^{g_i\rightarrow g_j}=(xy)^{g_i\rightarrow g_j}$ in eq. \ref{pij}; the traditional Directed Configuration Model (DCM), on the other hand, is obtainable by posing $x_i^{g_i\rightarrow g_j}y_j^{g_i\rightarrow g_j}=x_iy_j$ in the same equation. Upon eliminating the parameters dependence on nodes, $x_i^{g_i\rightarrow g_j}y_j^{g_i\rightarrow g_j}=xy$ and the Directed Random Graph Model (DRG) is finally obtained.

Interestingly, the \emph{directed degree-corrected SBM} (ddc-SBM) can be recovered by decoupling the parameters dependence on node-specific quantities from their group membership, i.e. by posing $x_i^{g_i\rightarrow g_j}y_j^{g_i\rightarrow g_j}=x_iy_j\chi^{g_i\rightarrow g_j}$.\\

When analysing directed networks, however, a non-trivial piece of information to be taken into account is represented by reciprocity \cite{Garlaschelli2006}. For this reason, a second class of null models, i.e. the one including the so-called \emph{reciprocity-informed null models}, is considered as well. Null models in this class are defined by constraints encoding the (non) reciprocal degree sequences, beside the usual nodes membership. In the most general case, the constraints defining such models can be written as

\begin{eqnarray}
k_i^{\xrightarrow{rs}}&=&\delta_{g_ir}\sum_{j(\neq i)}\delta_{g_js}a_{ij}^\rightarrow,\:\forall\:i,r,s\\
k_i^{\xleftarrow{rs}}&=&\delta_{g_ir}\sum_{j(\neq i)}\delta_{g_js}a_{ij}^\leftarrow,\:\forall\:i,r,s \\
k_i^{\xleftrightarrow{rs}}&=&\delta_{g_ir}\sum_{j(\neq i)}\delta_{g_js}a_{ij}^\leftrightarrow,\:\forall\:i,r,s.
\end{eqnarray}
with $a_{ij}^{\rightarrow}=a_{ij}(1-a_{ji})$, $a_{ij}^{\leftarrow}=a_{ji}(1-a_{ij})$ and $a_{ij}^{\leftrightarrow}=a_{ij}a_{ji}$ \cite{Garlaschelli2006} and $k_i^{\xleftrightarrow{rs}}$ indicating the contribution to the reciprocal degree of node $i$ (belonging to block $r$) coming from block $s$. All models in this second class induce a probability for the network $\mathbf{A}$ reading 

\begin{equation}
P(\mathbf{A})=\prod_{i<j}(p_{ij}^\rightarrow)^{a_{ij}^\rightarrow}(p_{ij}^\leftarrow)^{a_{ij}^\leftarrow}(p_{ij}^\leftrightarrow)^{a_{ij}^\leftrightarrow}(p_{ij}^\nleftrightarrow)^{a_{ij}^\nleftrightarrow};
\end{equation}
as before, different null models induce different functional forms for the probability coefficients $\{p_{ij}^\rightarrow\}$, $\{p_{ij}^\leftarrow\}$, $\{p_{ij}^\leftrightarrow\}$, $\{p_{ij}^\nleftrightarrow\}$: more explicitly, while the Reciprocal Configuration Model (RCM) is defined by the set of equations

\begin{eqnarray}
p_{ij}^\rightarrow&=&\frac{x_i y_j}{1+x_i y_j+y_j x_i+z_i z_j},\\
p_{ij}^\leftarrow&=&\frac{x_j y_i}{1+x_i y_j+y_i x_j+z_i z_j},\\
p_{ij}^\leftrightarrow&=&\frac{z_i z_j}{1+x_i y_j+y_i x_j+z_i z_j}
\end{eqnarray}
its block-wise counterpart, i.e. the Block Reciprocal Configuration Model (BRCM), is defined by the block-specific version of the coefficients above (see Appendix A for more details).\\

Models in both classes are \emph{parametric}: a recipe is, then, needed to estimate the parameters appearing in their definition. To this aim, the likelihood-maximization principle can be invoked, the likelihood function associated with $P(\mathbf{A})$ reading $\mathcal{L}(\mathbf{A})=\ln P(\mathbf{A})$. Notably, the evidence that each null model we consider in this paper treats different nodes pairs as independent allows us to write the likelihood for block models in a block-wise form, i.e. as $\mathcal{L}(\mathbf{A})=\ln\left[\prod_b P(\mathbf{A}^{(b)})\right]=\sum_b\ln P(\mathbf{A}^{(b)})$ with $b$ indexing the different modules (e.g. $b=S,\:I,\:O\dots$ in the case of bow-tie structures).

\subsection{Model selection criteria}

Although rising the number of parameters to better reproduce empirical patterns is tempting, the risk of overfitting should be, nevertheless, avoided. A criterion to identify the best model out of a basket of possible ones is, thus, needed. In what follows, we will adopt the Akaike Information Criterion (AIC hereafter)

\begin{equation}\label{aic}
\text {AIC}=-2\mathcal{L}(\mathbf{A})+2K+\frac{2K(K+1)}{n-K-1}
\end{equation}
and the Bayesian Information Criterion (BIC hereafter)

\begin{equation}\label{bic}
\text{BIC}=-2\mathcal{L}(\mathbf{A})+K\ln n
\end{equation}
whose first addendum is, in both cases, proportional to the likelihood of the null model under analysis, $K$ is the number of parameters defining the model and $n$ is the sample size (set, as usual, at $N(N-1)$). Both AIC and BIC are minimum for the best explanatory model in the basket \cite{Burnham2011}.

In order to make eqs. \ref{aic} and \ref{bic} more explicit, let us call $B$ the number of blocks our network can be divided into (i.e. the \emph{diagonal} blocks of the matrix $\mathbf{A}$). While the Directed Random Graph (DRG) is defined by just one parameter, $K_{DRG}=1$, the Stochastic Block Model (SBM) is defined by $K_{SBM}=B^2$ parameter (as can be verified upon inspecting definitions \ref{cp} and \ref{bt}).

Specifying the degree sequences leads to further rise the number of parameters: the Directed Configuration Model (DCM) is, in fact, defined by $K_{DCM}=2N$, the directed degree-corrected Stochastic Block Model (ddc-SBM) is defined by $K_{ddc-SBM}=2N+B^2$ and the Block Configuration Model (BCM) is defined by $K_{BCM}=2NB$ (each node, in fact, ``needs'' two parameters per block).

Accounting also for the information provided by the reciprocity requires a number of parameters to be specified that is $K_{RCM}=3N$ for the Reciprocal Configuration Model (RCM) and $K_{BRCM}=3NB$ for the Block Reciprocal Configuration Model (BRCM - each node, in fact, ``needs'' three parameters per block).

\begin{figure}[t!]
\begin{center}
\includegraphics[width=0.42\textwidth]{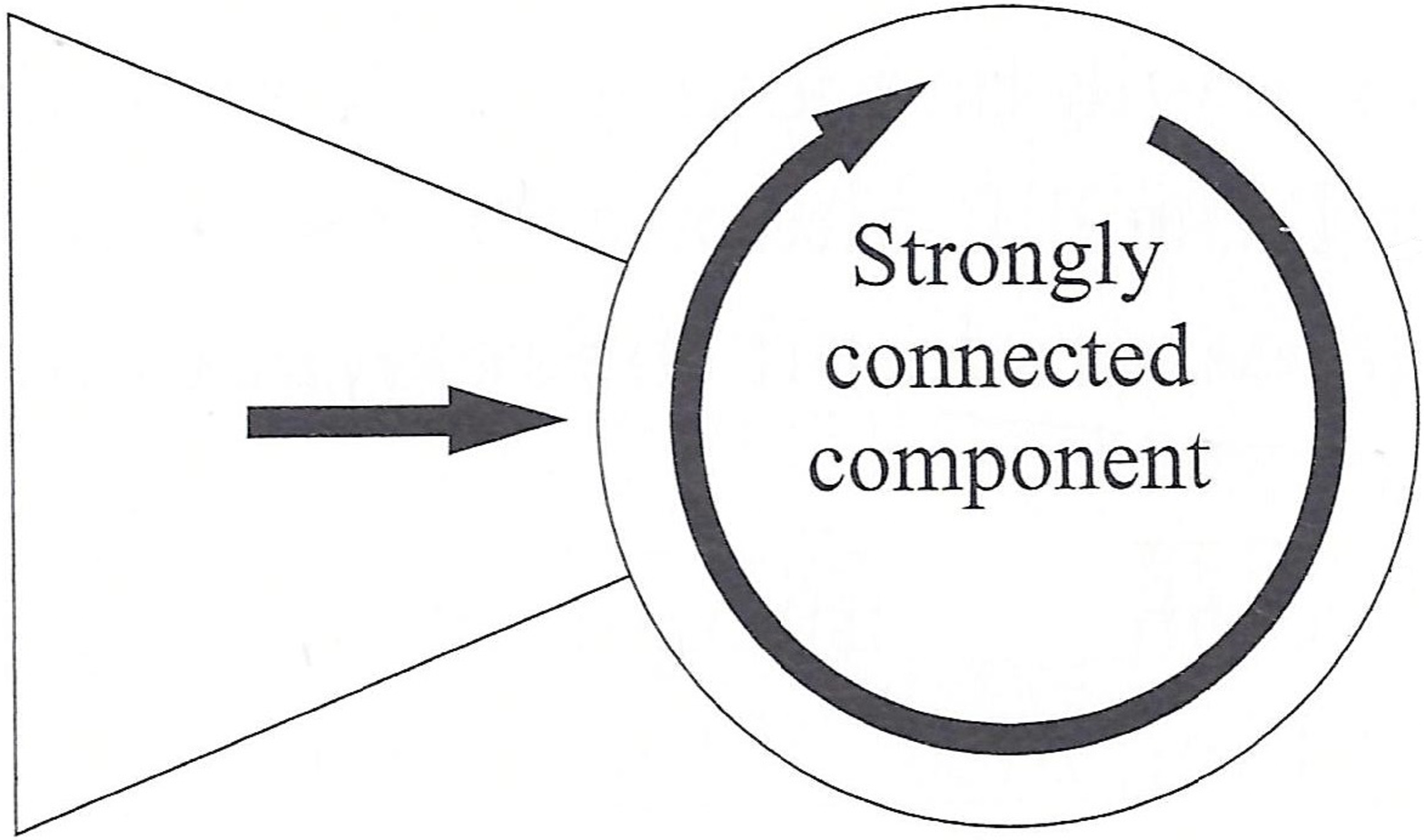}
\includegraphics[width=0.49\textwidth]{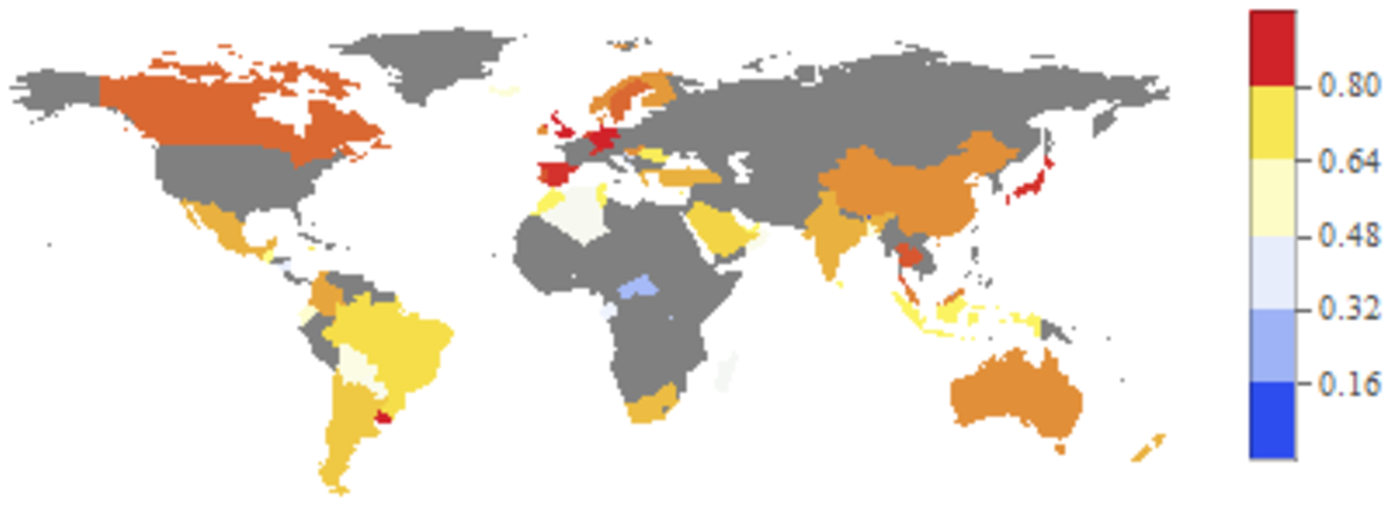}
\includegraphics[width=0.49\textwidth]{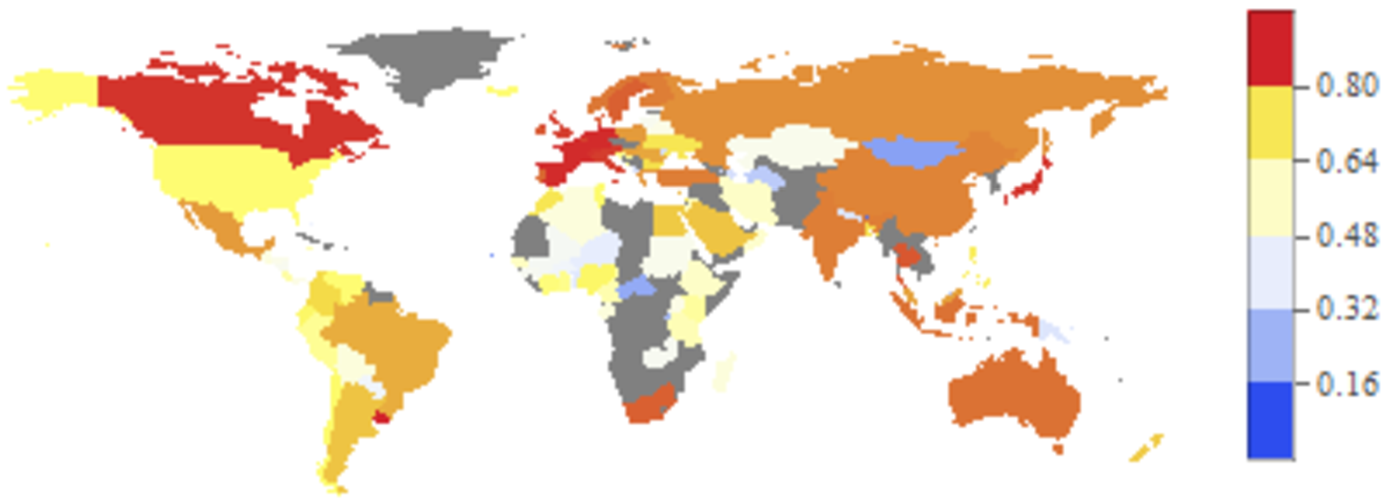}
\includegraphics[width=0.49\textwidth]{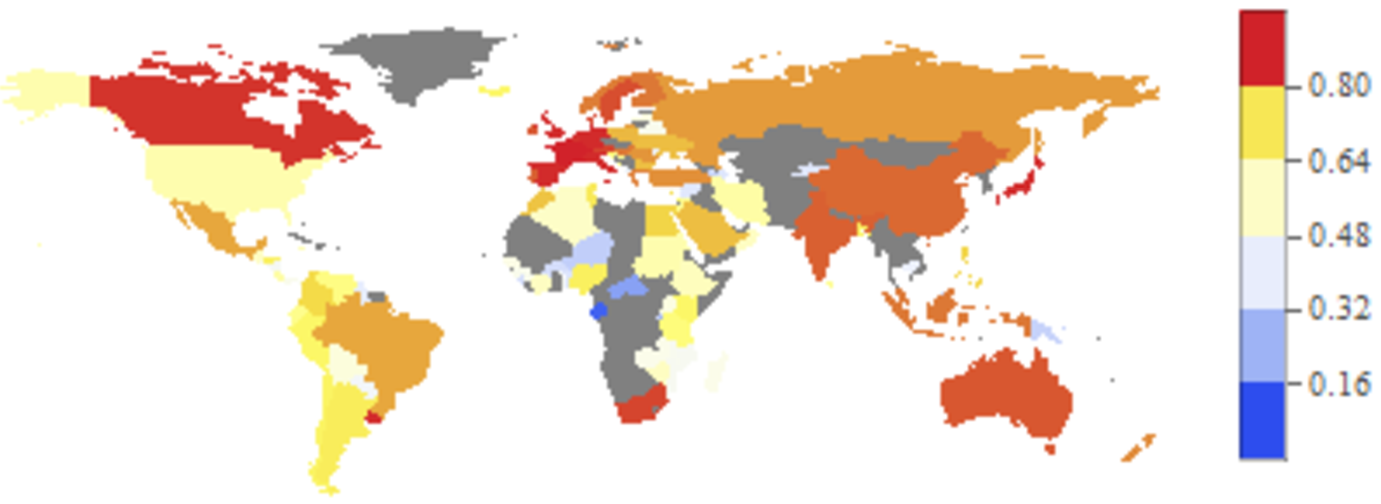}
\caption{Top panel: the WTW bow-tie structure, composed by the SCC and the IN-component only. The panels below show the countries belonging to the SCC (in colors) and the countries belonging to the IN-component (in gray) in 1993, 1998 and 2002, respectively. Countries belonging to the SCC keep rising their reciprocated degree (see also fig. \ref{fig2}); richest world countries (Canada, Europe, Japan - in dark red) are always characterized by the largest values of reciprocated degree.\label{fig1}}
\end{center}
\end{figure}

\begin{figure*}[t!]
\begin{center}
\includegraphics[width=0.49\textwidth]{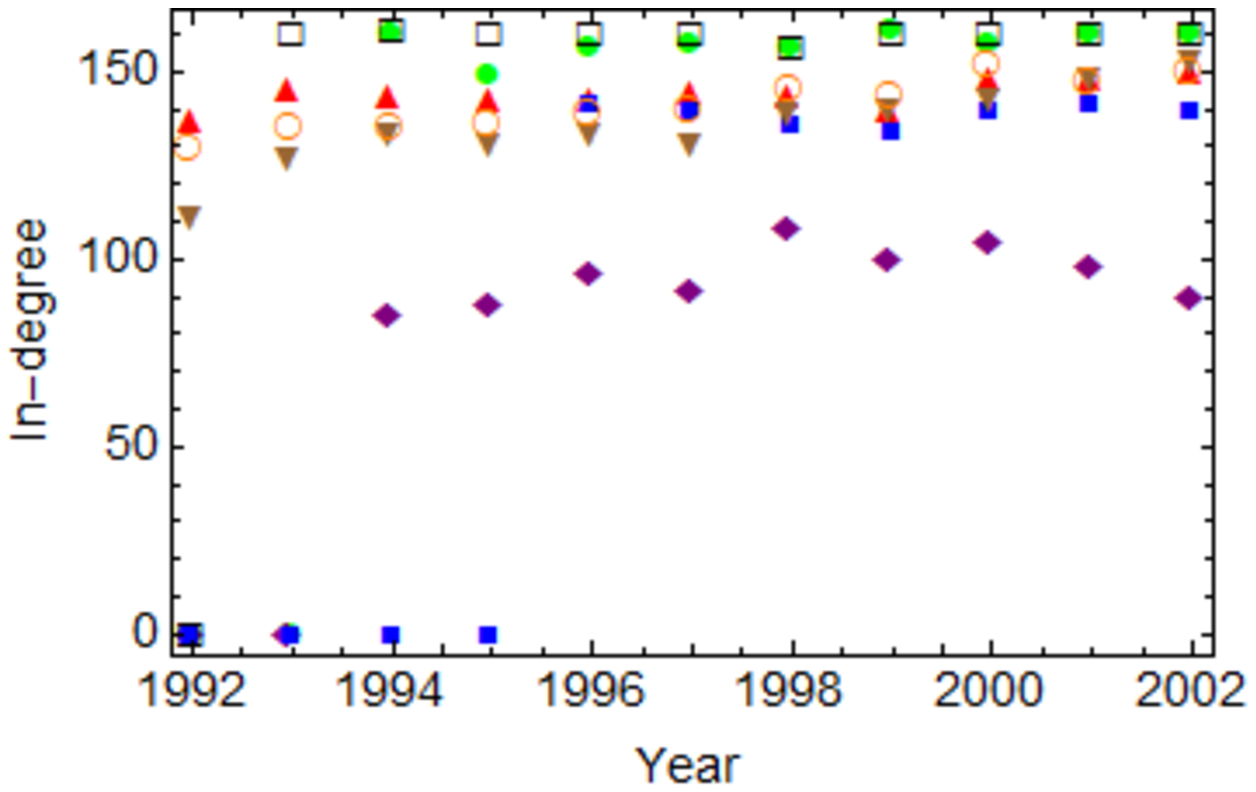}
\includegraphics[width=0.49\textwidth]{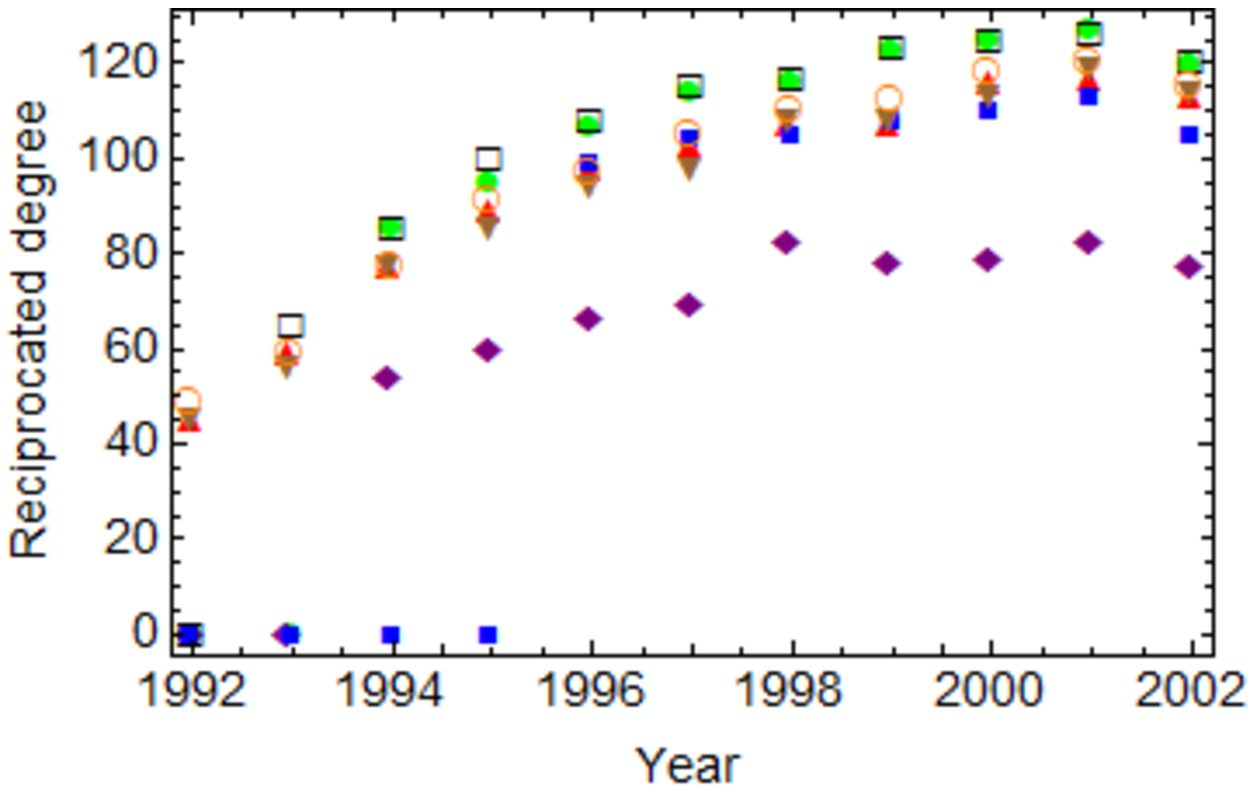}
\caption{Dynamics of the in-degree (defined as $h_i=\sum_{j(\neq i)}a_{ji}$) and of the reciprocated degree (defined as $k_i^\leftrightarrow=\sum_{j(\neq i)}a_{ij}a_{ji}$) of a sample of countries (Italy, in green; Japan, in black; China, in red; Russia, in blue; India, in brown; USA, in purple; Australia, in orange): while the in-degree remains rather stable across time, the value of the reciprocated degree keeps rising once the country has joined the SCC. Such a dynamics can be interpreted as a signal of ongoing integration \cite{Barigozzi2010}.\label{fig2}}
\end{center}
\end{figure*}

The model selection framework based upon the two information criteria above allows the probability that a given model $m$ is the best approximating model to be calculated as well, via the so-called \emph{AIC weights} and \emph{BIC weights}, defined as

\begin{equation}
w_m=\frac{e^{-\Delta_m/2}}{\sum_me^{-\Delta_m/2}}
\end{equation}
with $\Delta_m=\text{AIC}_m-\min\{\text{AIC}_m\}_m$ and $\Delta_m=\text{BIC}_m-\min\{\text{BIC}_m\}_m$, respectively.

\section{Results\label{results}}

\paragraph*{The World Trade Web.} Although the WTW has been deeply studied throughout the years \cite{Serrano2003,Fronczak2012,Fagiolo2012,Mastrandrea2014b}, the analysis of its mesoscale organization has received far less attention \cite{Barigozzi2010,Torreggiani2017}. Interestingly, checking for the applicability of the bow-tie definition provided above, the WTW appears as being partitioned into a SCC and an IN-component only, the OUT-component being completely missing (see fig. \ref{fig1}). According to the algebraic representation introduced at the beginning of the paper, the WTW mesoscale structure is represented by the following adjacency matrix

\begin{equation}\label{wtwbt}
\mathbf{A}^{\text{WTW}}=\left( \begin{array}{cc}
\mathbf{A}^{I} & \mathbf{A}^\rangle\\
\mathbf{0} & \mathbf{A}^{S}\\
\end{array} \right)
\end{equation}
with $\mathbf{A}^{I}=\mathbf{0}$ throughout our temporal interval. This implies that the nodes belonging to the IN-component do not establish internal relationships, their links pointing towards the SCC nodes only (via the $\mathbf{A}^\rangle$ block). Interestingly, the percentage of nodes belonging to the SCC steadily increases with time: from the 32\% in 1992 to almost the 75\% in 2002. Since the total number of nodes does not vary across the considered temporal interval, the IN-component shrinks accordingly. These results refine the picture drawn in \cite{Barigozzi2010}, where only the largest connected component was considered.

From a macroeconomic point of view, the increasing number of nodes within the SCC may evidence a sort of ongoing globalization process \cite{Barigozzi2010}. It is interesting to notice that the inclusion of (whole subsets of) countries within the SCC seems to be related to the existence of trade agreements. Examples are provided by Commonwealth nations - all of which are part of the SCC since 1993 - European nations (EU as a whole joined the SCC in 1994, the same year of the EEA agreement) and the case of USA (NAFTA entered into force in 1994 as well). From a purely topological perspective, an interesting dynamics takes place: as shown in fig. \ref{fig2}, the reciprocal degree of nodes belonging to the SCC keeps rising. Since all nodes are characterized by a rather stable in-degree value, this finding points out the tendency of such countries to reciprocate previously-established connections by creating new out-going links (i.e. to consolidate existing trade relationships). Besides, such a dynamics suggests that the large number of paths within the SCC may be due to the large value of reciprocity characterizing it.\\

Let us now analyse what kind of topological information is actually needed in order to explain the mesoscale WTW structure. To this aim, let us sum up the observations about the empirical structure of the WTW by imagining a \emph{densely-connected}, \emph{highly-reciprocated} SCC - $c(\mathbf{A}^{S})\simeq r(\mathbf{A}^{S})\simeq 0.8$ - throughout our temporal interval.

\begin{figure*}[t!]
\begin{center}
\includegraphics[width=\textwidth]{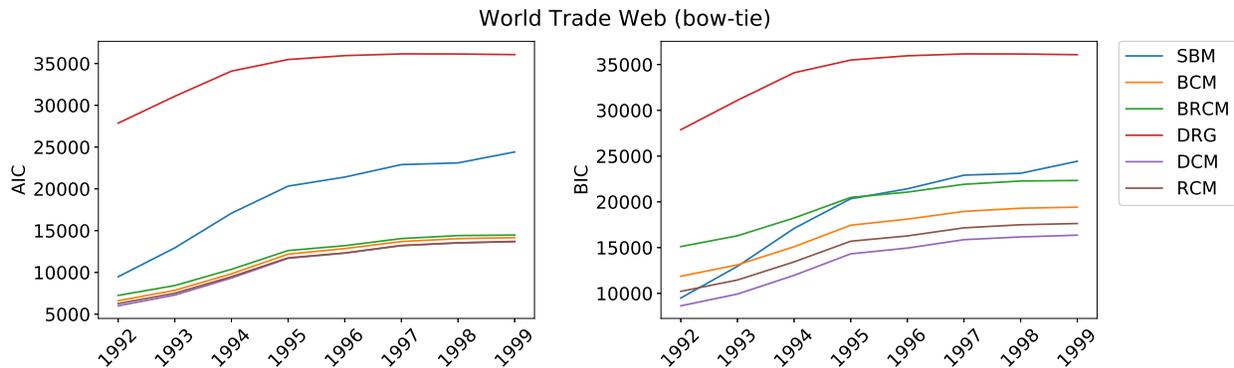}
\caption{Evolution of the AIC and BIC values for the WTW across the years 1992-2002: while the SBM (blue trend) must be preferred to the traditional DRG (being the network composed by parts with different link densities), heterogeneous benchmarks are, generally speaking, to be preferred. Although the DCM and the RCM are characterized by very similar AIC values, AIC and BIC weights let always the DCM win. The ddc-SBM experiences convergence problems throughout the entire temporal period.\label{fig3}}
\end{center}
\end{figure*}

The need of considering a block model becomes evident when comparing the homogeneous benchmark provided by the DRG with its block-wise counterpart, i.e. the SBM (see fig. \ref{fig3}). The SBM outperforms the DRG since the network is ``composed'' by modules characterized by very different link-densities that cannot be reproduced by tuning just one, global parameter: in fact, $c(\mathbf{A}^{S})\in[0.75, 0.9]$ and $c(\mathbf{A}^{I})=0$.

Generally speaking, however, benchmarks encoding the degree heterogeneity are to be preferred. Interestingly, (both) non-block models outperform block models, indicating that specifying additional information to the one encoded into local properties is indeed unnecessary. This is not surprising, however, when considering that the nodes belonging to the IN-component have zero in-degrees. The latter, in fact, are exactly reproduced by both the DCM and the RCM: the ``peripherical'' part of the network under analysis is, thus, automatically explained by a simpler kind of statistics with no need to invoke any \emph{a priori} partition.

Let us now compare our degree-informed models over the $\mathbf{A}^\rangle$ and $\mathbf{A}^{S}$ subgraphs. For what concerns the former, the information carried by reciprocity is encoded into the degree sequences: the result $\mathcal{L}(\mathbf{A}^\rangle)_{BCM}=\mathcal{L}(\mathbf{A}^\rangle)_{BRCM}$ is, in fact, rooted into the observation that the links from the IN-component to the SCC are not reciprocated.

The same consideration, together with the observation that the large $r(\mathbf{A}^S)$ value is due to reciprocal connections established between nodes \emph{within} the SCC, leads to the result $\mathcal{L}(\mathbf{A}^\rangle)_{BCM}=\mathcal{L}(\mathbf{A}^\rangle)_{BRCM}\simeq\mathcal{L}(\mathbf{A}^\rangle)_{RCM}$; similarly, $\mathcal{L}(\mathbf{A}^{S})_{BRCM}\simeq\mathcal{L}(\mathbf{A}^{S})_{RCM}$. As a consequence, being the two likelihood values (overall) very similar, the model with a larger number of parameters is more ``penalized'' (i.e. $\text{AIC}_{BRCM}>\text{AIC}_{RCM}$).

On the other hand, comparing the BCM and the DCM on the SCC leads to the conclusion that, as the latter enlarges, $\mathcal{L}(\mathbf{A}^{S})_{BCM}\simeq\mathcal{L}(\mathbf{A}^{S})_{DCM}$, since the largest contribution to the nodes degrees comes from the connections established with other nodes within the SCC itself.

Apparently, thus, two non-block models compete, i.e. the DCM and the RCM (see fig. \ref{fig2}). However, the computation of the AIC and BIC weights for each model $m$ in our basket reveals that the DCM always wins. The explanation of this result may lie in the evidence that the WTW reciprocity is actually compatible with the DCM prediction, as the computation of the index $\rho=\frac{r-\langle r\rangle}{1-r}$ reveals (it amounts at $\simeq 0.05$ throughout our time interval) \cite{Garlaschelli2004}. In other words, the seemingly peculiar mesoscale structure of the WTW is, to a good extent, reproduced by just specifying local constraints as the in- and out- degree sequences.\\

\paragraph*{The Dutch Interbank Network.} According to the axiomatic model in \cite{Craig2010}, the DIN has been described as characterized by a well-defined core-periphery structure \cite{Lelyveld2012}. However, as it has been pointed out elsewhere \cite{Squartini2013}, such a mesoscale organization is compatible with the predictions provided either by the DCM or the RCM, depending on the topological quantity inspected.

\begin{figure}[t!]
\begin{center}
\includegraphics[width=0.49\textwidth]{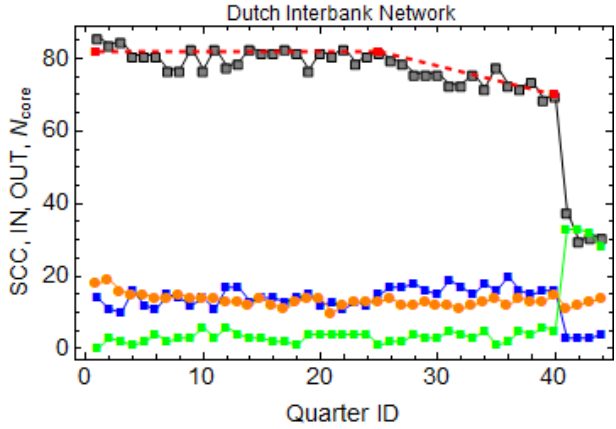}
\caption{Evolution of the DIN bow-tie structure (the SCC is shown in gray, the IN-component is shown in blue and the OUT-component is shown in green). The crisis period (last four points) is signalled by a sharp decrease of the SCC and IN- components size (and a corresponding increase of the OUT-component size). The size of the SCC, however, starts shrinking in 2004Q1 (deviating from the approximately costant trend observed since 1998Q1), seemingly constituting an additional, early-warning signal of the upcoming crisis. On the other hand, the DIN core (shown in orange) doesn't undergo any significant variation throughout the whole temporal interval.\label{fig4}}
\end{center}
\end{figure}

Notably, the DIN is also characterized by a certain degree of bow-tieness, given the presence of an SCC, an IN-component and, differently from the WTW, also a non-vanishing OUT-component: both the $\mathbf{A}^{I}$ and the $\mathbf{A}^{O}$ blocks, however, are empty, and nodes belonging to the IN- and OUT- components are not directly linked with each other (but only via the SCC nodes). From a purely empirical point of view, the evolution of the DIN bow-tie structure is much more informative than the evolution of its core-periphery structure: as fig. \ref{fig4} shows, while the size of the DIN SCC, in 2008, reduces to more than half its pre-crisis value - thus providing an additional, structural indicator of it - the number of nodes belonging to the core shows no significant variations across the same period. Very interestingly, however, the SCC starts shrinking well before 2008, a dynamics seemingly constituting an additional early-warning signal of the upcoming, topological change affecting the DIN. The IN-component, in turn, shrinks as well, while the OUT-component enlarges.\\

\begin{figure*}[t!]
\begin{center}
\includegraphics[width=\textwidth]{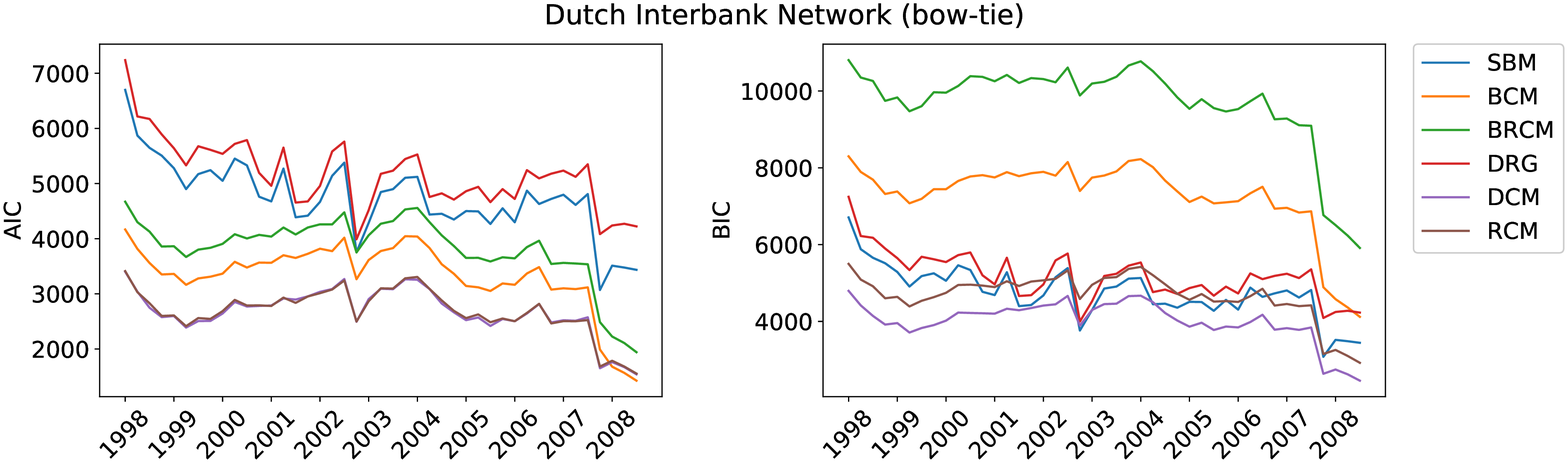}
\includegraphics[width=\textwidth]{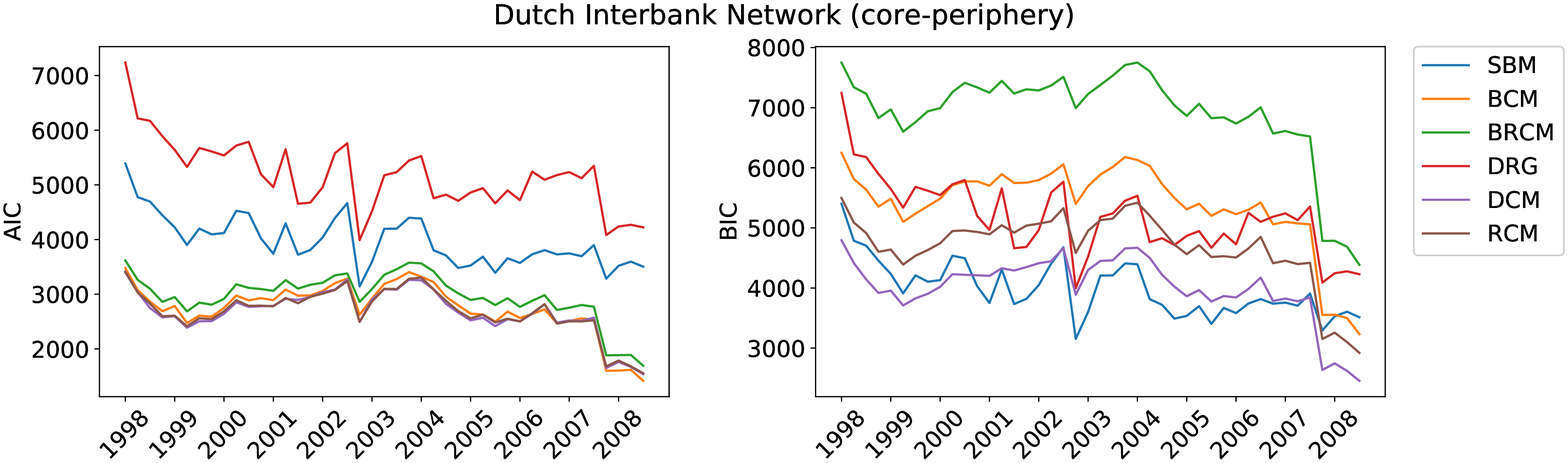}
\caption{Evolution of the AIC and BIC values for the DIN across the quarters 1998Q1-2008Q4: while the SBM (blue trend) must be preferred to the traditional DRG (being the network composed by parts with different link densities), heterogeneous benchmarks are, generally speaking, to be preferred. Although the DCM wins in the vast majority of cases (both for the bow-tie and the core-periphery mesoscale structures), quarters exist where the DCM and the RCM compete; BIC, on the other hand, lets the SBM win sometimes, when analysing the DIN core-periphery structure. Red crosses mark the periods where the ddc-SBM converges.\label{fig5}}
\end{center}
\end{figure*}

In order to individuate the null model encoding the right amount of topological information to explain the DIN bow-tie structure, let us notice that its SCC can be imagined as a \emph{weakly-connected}, \emph{weakly-reciprocated} subgraph ($c(\mathbf{A}^{S})\lesssim 0.1$ and $r(\mathbf{A}^{S})\simeq 0.3$, except in 2008 where the SCC reciprocity drops to $\simeq 0.15$). More precisely, $c(\mathbf{A}^{S})\gtrsim c(\mathbf{A})\ll c(\mathbf{A}^\bullet)$, i.e. while the SCC connectance basically coincides with the one of the whole network, the core is much denser, an empirical observation that explains why the SBM provides a better explanation of the core-periphery structure - see fig. \ref{fig5}. Conversely, the AIC and BIC values for the SBM and the DRG are closer when considering the bow-tie structure).

Generally speaking, however, models accounting for the degree heterogeneity are to be preferred. As for the WTW, zero in-degrees and zero out-degrees are exactly reproduced by non-block models models like the DCM and the RCM. On top of this, the low reciprocity value of the DIN (amounting at $\lesssim 0.3$) allows us to imagine it playing a minor role in determining the nodes degrees. As a consequence, the DCM and the RCM can be interpreted as different ways to rewrite the same (configuration) model. More quantitatively, $\mathcal{L}(\mathbf{A})_{RCM}\gtrsim\mathcal{L}(\mathbf{A})_{DCM}$. 

Deviations from this idealized picture, however, exist. This is particularly evident when analysing the $\mathbf{A}^{S}$ block, to fully understand which reciprocity indeed plays a role (in fact, $\mathcal{L}(\mathbf{A}^{S})_{BRCM}>\mathcal{L}(\mathbf{A}^{S})_{BCM}$); when considering the ``peripherical'' blocks, instead, one concludes that $\mathcal{L}(\mathbf{A}^\rangle)_{BRCM}\simeq\mathcal{L}(\mathbf{A}^\rangle)_{BCM}$, $\mathcal{L}(\mathbf{A}^\rangle)_{RCM}\gtrsim\mathcal{L}(\mathbf{A}^\rangle)_{DCM}$ and $\mathcal{L}(\mathbf{A}^{\rangle\rangle})_{BRCM}\simeq\mathcal{L}(\mathbf{A}^{\rangle\rangle})_{BCM}$, $\mathcal{L}(\mathbf{A}^{\rangle\rangle})_{RCM}\gtrsim\mathcal{L}(\mathbf{A}^{\rangle\rangle})_{DCM}$ (since the links from the IN-component to the SCC and from the SCC to the OUT-component are not reciprocated).

Consistently, AIC and BIC weights let the DCM win in the vast majority of cases, although in some periods the DCM and the RCM compete. Overall, this is valid when considering the DIN core-periphery structure too.

\section{Discussion\label{discussion}}

The WTW and the DIN represent two real-world systems characterized by (apparently) non-trivial mesoscale structures: while the first one is characterized by a (partial) bow-tie organization, in the second one the bow-tie partition co-exists with a core-periphery partition. Both kinds of mesoscale structures are characterized by interacting blocks whose internal topology is commonly believed to be determined by a non-trivial interplay between nodes connectivity and the reciprocity of connections. It is, thus, interesting to ask ourselves the extent to which such structures are, instead, accounted for by purely local information.

Remarkably, what our analysis points out is that specifying the degree sequences is often enough to reproduce these mesoscale structures, thus suggesting that the observed modules emerge as a consequence of local connectivity patterns between nodes: for example, the absence of incoming/outgoing connections for a set of nodes naturally leads them to be identified as an IN-/OUT-component.

Differences between systems, naturally, exist. Let us notice that, contrarily to what observed in the WTW case, AIC and BIC provide different answers to the question concerning the performance of block models in explaining the DIN core-periphery structure: while the Akaike criterion ranks the BCM first, the Bayesian criterion assigns the highest score to the SBM in the vast majority of temporal snapshots. If, on the one hand, this saves the role potentially played by blocks, on the other it points out that the large difference between the connectivity values of the core and the periphery \cite{Squartini2013} provides - by itself - an effective explanation of this mesoscale organization.

A second comment about the DIN concerns the observation that, when considering the core-periphery structure, the AIC values of block models overlap with the AIC values of the simpler models to a larger extent (see fig. \ref{fig5}): this may be a consequence of the fact that the core-periphery partition is, in some sense, less ``neat'' than the bow-tie one (the requirement that nodes belonging to either the IN- or OUT- components have zero in- or out-degree represents a quite strong constraint); only apparently, however, the core-periphery organization seems to require additional information to be explained, as the explicit calculation of the Akaike weights confirms.

A third comment concerns reciprocity: although it plays a role in the definition of the ``core'' parts (i.e. the SCC and the properly-defined core), its explanatory power is much more limited than expected: as a result, the degree sequence seems to encode all relevant information to reproduce the mesoscale structures considered in the present paper, thus questioning the role supposedly played by some kind of higher-level information - e.g. a partition into blocks - to explain them.

\section*{Appendix A}

Generally speaking, all null models considered in this paper can be recovered within the Exponential Random Graphs (ERG) framework. Following \cite{Squartini2011}, a canonical ensemble $\mathcal{G}$ of adjacency matrices must be considered, in order to maximize Shannon entropy $S=-\sum_{\mathbf{A}\in\mathcal{G}}P(\mathbf{A})\ln P(\mathbf{A})$ under a given set of constraints $\vec{C}(\mathbf{A})$ \cite{Squartini2011}. A probability coefficient $P(\mathbf{A})$ is, then, assigned to every adjacency matrix in the esemble. The result of the aforementioned constrained-optimization problem is the well-known exponential distribution: $P(\mathbf{A}|\vec{\theta})=e^{-H(\mathbf{A},\vec{\theta})}/Z(\vec{\theta})$ with the hamiltonian $H(\mathbf{A},\vec{\theta})=\vec{\theta}\cdot\vec{C}(\mathbf{A})$ summing up the imposed set of constraints and $Z(\vec{\theta})=\sum_{\mathbf{A}\in\mathcal{G}}e^{-H(\mathbf{A},\vec{\theta})}$ being the normalization.

\subsection*{Degree-informed null models}

All degree-informed null models can be recovered as particular cases of the following hamiltonian

\begin{eqnarray}\label{gh}
H&=&\sum_{i\neq j}(\alpha_i^{g_i\rightarrow g_j}+\beta_j^{g_i\rightarrow g_j})a_{ij}=\nonumber\\
&=&\sum_{i\neq j}\sum_r\sum_s\delta_{g_ir}\delta_{g_js}(\alpha_i^{r\rightarrow s}+\beta_j^{r\rightarrow s})a_{ij}\hspace{5mm}
\end{eqnarray}
defined by constraints encoding the dependence on block-specific, local quantities, in the most general case.\\

\paragraph*{Block Configuration Model (BCM).} The BCM is defined by the probability coefficients introduced in eq. \ref{pij}, i.e.

\begin{equation}
p_{ij}=\frac{x_i^{g_i\rightarrow g_j}y_j^{g_i\rightarrow g_j}}{1+x_i^{g_i\rightarrow g_j}y_j^{g_i\rightarrow g_j}}
\end{equation}
(where $x_i=e^{-\alpha_i^{g_i\rightarrow g_j}}$ and $y_i=e^{-\beta_i^{g_i\rightarrow g_j}}$), to be numerically determined by solving the likelihood equations

\begin{equation}\label{bicm}
\left\{
\begin{array}{l}
k_i^{r\rightarrow s}=\langle k_i^{r\rightarrow s}\rangle,\:\forall\:i,r,s\\
h_i^{s\rightarrow r}=\langle h_i^{s\rightarrow r}\rangle,\:\forall\:i,r,s
\end{array}
\right.
\end{equation}
with $\langle k_i^{r\rightarrow s}\rangle=\delta_{g_ir}\sum_{j(\neq i)}\delta_{g_js}p_{ij}$ and $\langle h_i^{s\rightarrow r}\rangle=\delta_{g_ir}\sum_{j(\neq i)}\delta_{g_js}p_{ji}$. The BCM extends the results in \cite{Karrer2010,Fronczak2013} to the directed case.\\

\paragraph*{Directed Degree-Corrected SBM (ddc-SBM).} Interestingly, upon identifying $\alpha_i^{g_i\rightarrow g_j}\equiv\alpha_i+\frac{w^{g_i\rightarrow g_j}}{2}$ and $\beta_j^{g_i\rightarrow g_j}\equiv\beta_j+\frac{w^{g_i\rightarrow g_j}}{2}$ the \emph{directed degree-corrected SBM} (ddc-SBM) is recovered. Upon retaining all multipliers in eq. \ref{gh} and defining $x_i\equiv e^{-\alpha_i}$, $y_i\equiv e^{-\beta_i}$ and $\chi^{g_i\rightarrow g_j}\equiv e^{-w^{g_i\rightarrow g_j}}$, one finds

\begin{equation}\label{dcsbm}
p_{ij}=\frac{x_iy_j\chi^{g_i\rightarrow g_j}}{1+x_iy_j\chi^{g_i\rightarrow g_j}};
\end{equation}
although formally equivalent, expressions \ref{dcsbm} and \ref{pij} are not when coming to estimate the unknown parameters: eq. \ref{dcsbm} is, in fact, determined by solving the equations 

\begin{equation}
\left\{
\begin{array}{l}
k_i=\langle k_i\rangle,\:\forall\:i\\
h_i=\langle h_i\rangle,\:\forall\:i\\
L_{rs}=\langle L_{rs}\rangle,\:\forall\:r,s
\end{array}
\right.
\end{equation}
thus requiring less parameters than the BCM \cite{Reichardt2011}. The ddc-SBM generalizes the results in \cite{Karrer2010,Zhu2012} to the non-sparse case.\\

\paragraph*{Directed Configuration Model (DCM).} The DCM is obtained by posing $\alpha_i^{g_i\rightarrow g_j}=\alpha_i$ and $\beta_j^{g_i\rightarrow g_j}=\beta_j$ in eq. \ref{gh}. Upon defining $x_i\equiv e^{-\alpha_i}$ and $y_i\equiv e^{-\beta_i}$, the surviving multipliers induce probability coefficients reading

\begin{equation}
p_{ij}=\frac{x_iy_j}{1+x_iy_j}
\end{equation}
to be numerically determined by solving the likelihood equations

\begin{equation}
\left\{
\begin{array}{l}
k_i=\langle k_i\rangle,\:\forall\:i\\
h_i=\langle h_i\rangle,\:\forall\:i
\end{array}
\right.
\end{equation}
with the out- and in-degrees reading $k_i=\sum_{j(\neq i)}a_{ij}$ and $h_i=\sum_{j(\neq i)}a_{ji}$ respectively and $\langle k_i\rangle=\sum_{j(\neq i)}p_{ij}$, $\langle h_i\rangle=\sum_{j(\neq i)}p_{ji}$.\\

\paragraph*{Stochastic Block Model (SBM).} Notice that the directed version of the Stochastic Block Model (SBM) can be recovered as a special case of the BCM, by posing $\alpha_i^{g_i\rightarrow g_j}=\alpha^{g_i\rightarrow g_j}$ and $\beta_j^{g_i\rightarrow g_j}=\beta^{g_i\rightarrow g_j}$ in eq. \ref{gh} and solving the equations 

\begin{equation}
L_{rs}=\langle L_{rs}\rangle,\:\forall\:r,s
\end{equation}
with $L_{rs}=\sum_{i\neq j}\delta_{g_ir}\delta_{g_js}a_{ij}$ and $\langle L_{rs}\rangle=\sum_{i\neq j}\delta_{g_ir}\delta_{g_js}p_{ij}$.\\

\paragraph*{Directed Random Graph Model (DRG).} The DRG can be recovered as a particular case of the DCM, obtained by posing $\alpha_i\equiv\alpha$ and $\beta_j\equiv\beta$ in eq. \ref{gh}. The only coefficient $p_{ij}\equiv p$ is determined by solving the equation

\begin{equation}
L=\langle L\rangle
\end{equation}
with $L=\sum_{i\neq j}a_{ij}$ and $\langle L\rangle=\sum_{i\neq j}p$.

\subsection*{Reciprocity-informed null models}

\paragraph*{Reciprocal Configuration Model (RCM).} The RCM is defined by the following probability coefficients

\begin{eqnarray}
p_{ij}^\rightarrow&=&\frac{x_i y_j}{1+x_i y_j+y_j x_i+z_i z_j},\label{rcm}\\
p_{ij}^\leftarrow&=&\frac{x_j y_i}{1+x_i y_j+y_i x_j+z_i z_j},\label{rcm2}\\
p_{ij}^\leftrightarrow&=&\frac{z_i z_j}{1+x_i y_j+y_i x_j+z_i z_j}\label{rcm3}
\end{eqnarray}
to be numerically determined by solving the likelihood equations 

\begin{equation}
\left\{
\begin{array}{l}
k_i^\rightarrow=\langle k_i^\rightarrow\rangle,\:\forall\:i\\
k_i^\leftarrow=\langle k_i^\leftarrow\rangle,\:\forall\:i\\
k_i^\leftrightarrow=\langle k_i^\leftrightarrow\rangle,\:\forall\:i
\end{array}
\right.
\end{equation}
with $\langle k_i^\rightarrow\rangle=\sum_{j(\neq i)}p_{ij}^\rightarrow$, $\langle k_i^\leftarrow\rangle=\sum_{j(\neq i)}p_{ij}^\leftarrow$, $\langle k_i^\leftrightarrow\rangle=\sum_{j(\neq i)}p_{ij}^\leftrightarrow$.\\

\paragraph*{Block Reciprocal Configuration Model (BRCM).} The RCM can be re-defined in a block-wise fashion, by specifying the probability coefficients defined by eqs. \ref{rcm}, \ref{rcm2}, \ref{rcm3} for each block. A Block Reciprocal Configuration Model (BRCM), thus, remains naturally defined by the system of equations

\begin{equation}\label{brcm}
\left\{
\begin{array}{l}
k_i^{\xrightarrow{rs}}=\langle k_i^{\xrightarrow{rs}}\rangle,\:\forall\:i,r,s\\
k_i^{\xleftarrow{rs}}=\langle k_i^{\xleftarrow{rs}}\rangle,\:\forall\:i,r,s \\
k_i^{\xleftrightarrow{rs}}=\langle k_i^{\xleftrightarrow{rs}}\rangle,\:\forall\:i,r,s
\end{array}
\right.
\end{equation}
with obvious meaning of the symbols.

\section*{Appendix B}

Let us explicitly solve the BCM in the two, off-diagonal matrices $\mathbf{A}^\top$ and $\mathbf{A}^\bot$. In order to fix the formalism, let us suppose the two off-diagonal blocks $\mathbf{A}^\top$ and $\mathbf{A}^\bot$ to have dimensions $C\times P$ and $P\times C$, respectively. Analogously to the undirected case \cite{Saracco2015}, solving the DCM within the off-diagonal blocks of the matrix $\mathbf{A}$ induces the following probability coefficients

\begin{equation}
P(\mathbf{A}^\top)=\prod_c\prod_pp_{cp}^{a^\top_{cp}}(1-p_{cp})^{1-a^\top_{cp}}
\end{equation}
and
\begin{equation}
P(\mathbf{A}^\bot)=\prod_p\prod_cq_{pc}^{a^\bot_{pc}}(1-q_{pc})^{1-a^\bot_{pc}};
\end{equation}
the probability that a link from a core node $c$ to a periphery node $p$ exists is $p_{cp}\equiv \frac{x^\top_cy^\top_p}{1+x^\top_cy^\top_p}$ and the probability that a link from a periphery node $p$ to a core node $c$ exists is $q_{pc}\equiv \frac{x^\bot_py^\bot_c}{1+x^\bot_py^\bot_c}$. Consistently, the vector $\vec{x}=\{\vec{x}^{\top}_c,\:\vec{x}^{\bot}_p\}$ is coupled to the outgoing degrees, while the vector $\vec{y}=\{\vec{y}^{\bot}_c,\:\vec{y}^{\top}_p\}$ is coupled to the incoming degrees. 

The aforementioned probability coefficients are determined via the likelihood condition in \ref{bicm}. Let us notice that the out-degree of core nodes and the in-degree of periphery nodes are measured on the matrix $\mathbf{A}^\top$; the converse is true for the matrix $\mathbf{A}^\bot$. More quantitatively, upon indicating with $\{\vec{k},\vec{h}\}$ the core and periphery nodes degrees, one has

\begin{equation}
k_c^{out}=\sum_pa^\top_{cp},\:k_c^{in}=\sum_pa^\bot_{pc},\:\forall\:c
\end{equation}
and

\begin{equation}
h_p^{out}=\sum_ca^\bot_{pc},\:h_p^{in}=\sum_ca^\top_{cp},\:\forall\:p.
\end{equation}

The estimation step, thus, reads

\begin{equation}
\left\{ \begin{array}{ll}
k_c^{out}&=\:\:\:\sum_pp_{cp},\:\forall\:c,\\
h_p^{out}&=\:\:\:\sum_cq_{pc},\:\forall\:p,\\
k_c^{in}&=\:\:\:\sum_pq_{pc},\:\forall\:c,\\
h_p^{in}&=\:\:\:\sum_cp_{cp},\:\forall\:p.
\end{array}
\right.\\
\label{sys1}
\end{equation}

The SBM can be recovered by posing $p_{cp}\equiv p$ and $q_{cp}\equiv q$, to be estimated by solving

\begin{equation}
p=\frac{L^\top}{C\cdot P}=\frac{\sum_{c,p}a_{cp}^\top}{C\cdot P}\:\mbox{and}\:q=\frac{L^\bot}{C\cdot P}=\frac{\sum_{c,p}a_{cp}^\bot}{C\cdot P}
\end{equation}
with obvious meaning of the symbols.\\

Inserting the information about reciprocity into a bipartite null model leads to the following probability coefficient

\begin{equation}
P(\mathbf{B})=\prod_c\prod_p(p_{cp}^\rightarrow)^{a_{cp}^\rightarrow}(p_{cp}^\leftarrow)^{a_{cp}^\leftarrow}(p_{cp}^\leftrightarrow)^{a_{cp}^\leftrightarrow}(p_{cp}^\nleftrightarrow)^{a_{cp}^\nleftrightarrow}
\end{equation}
that ``mixes'' the information coming from the two biadjacency matrices $\mathbf{A}^\top$ and $\mathbf{A}^\bot$ (whence the choice of a different symbol, $\mathbf{B}$, to indicate the bipartite network as a whole). The new variables read $a_{cp}^{\rightarrow}=a_{cp}^{\top}(1-a_{pc}^{\bot})$, $a_{cp}^{\leftarrow}=a_{pc}^{\bot}(1-a_{cp}^{\top})$, $a_{cp}^{\leftrightarrow}=a_{cp}^{\top}a_{pc}^{\bot}$ and $a_{cp}^{\nleftrightarrow}=(1-a_{cp}^{\top})(1-a_{pc}^{\bot})$: while $a_{cp}^{\rightarrow}$ indicates that a non-reciprocated link is present from the core node $c$ to the periphery node $p$, $a_{cp}^{\leftarrow}$ indicates that a non-reciprocated link is present from the periphery node $p$ to the core node $c$; naturally, $a_{cp}^{\leftrightarrow}$ indicates that both links are present between nodes $c$ and $p$ and $a_{cp}^{\nleftrightarrow}$ indicates that no link is present between the same nodes.

\begin{figure*}[t!]
\begin{center}
\includegraphics[width=0.49\textwidth]{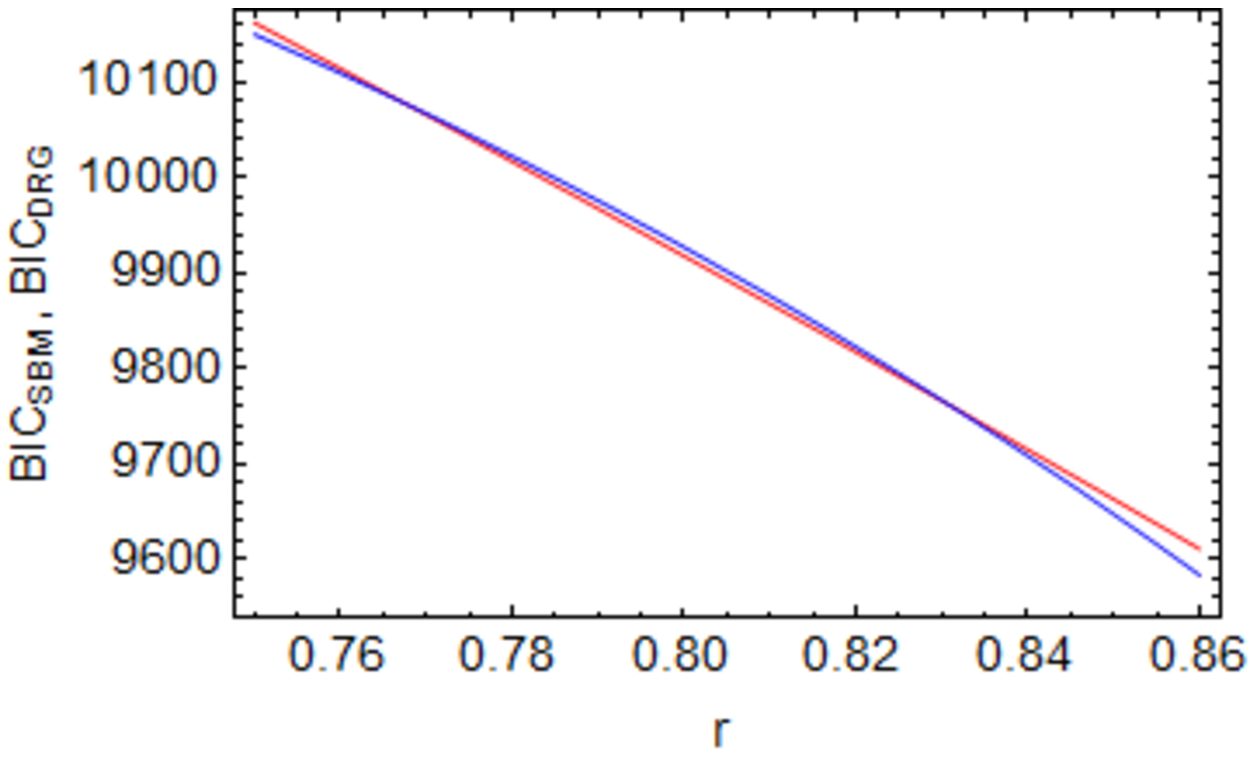}
\includegraphics[width=0.49\textwidth]{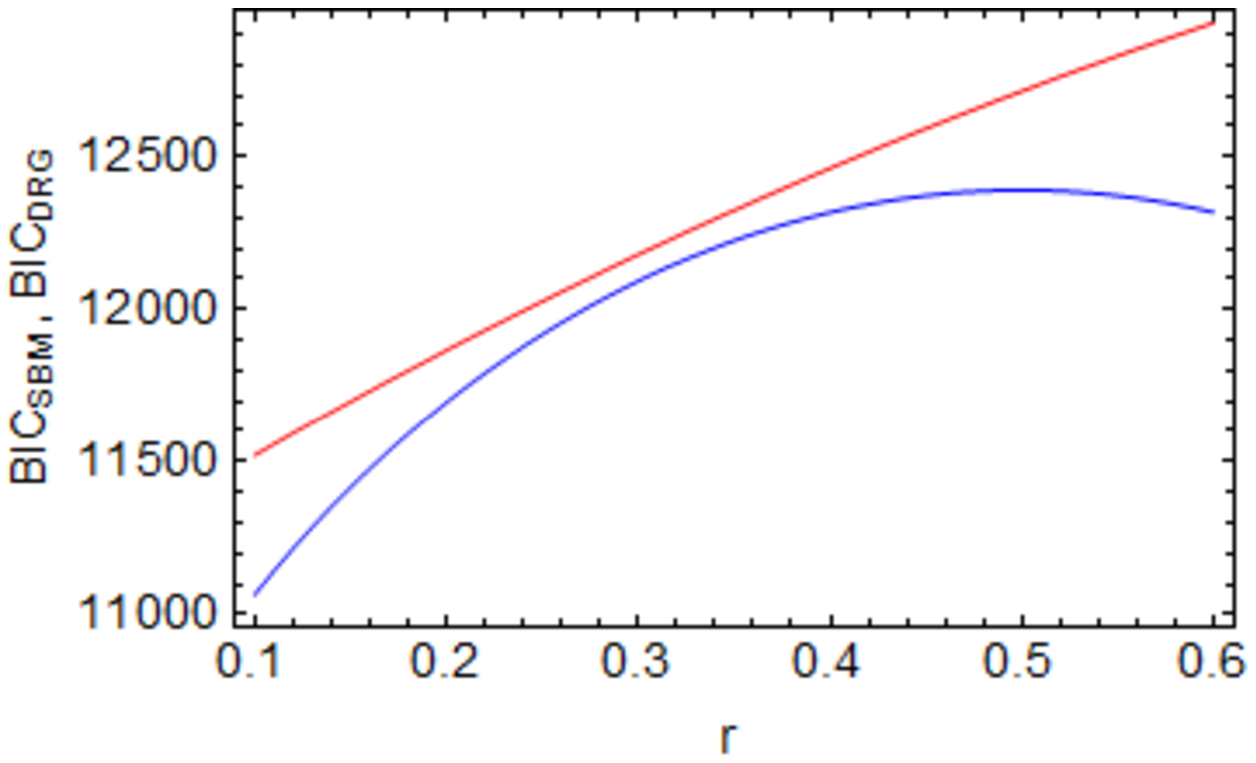}
\caption{Left panel: comparison between the numerical values of the BIC computed for the SBM (blue trend) and the DRG (red trend), on a bimodular network where the link density of the two communities ($N_1=10$, $N_2=90$) amounts at $p=q=0.8$. Notice that a region of values around $r=0.8$ exists where the SBM is penalized: only one global parameter is, in fact, enough to satisfactorily reproduce the network structure. As the link density of the off-diagonal blocks deviates from the value $r=0.8$ the network becomes more and more heterogeneous and specifying the modules is indeed rewarding. Right panel: comparison between the numerical values of the BIC computed for the SBM (blue trend) and the DRG (red trend), on a core-periphery network where the link density of the two blocks ($N_1=10$, $N_2=90$) amounts at $p^\bullet=0.8$ and $p^\circ=0.3$. Notice that no region of values exists where the DRG is to be preferred: the network, in fact, is so heterogeneous that only one global parameter is not enough to account for its structure.\label{fig6}}
\end{center}
\end{figure*}

The probability coefficients defining our bipartite, reciprocal model read

\begin{eqnarray}
p_{cp}^\rightarrow&=&\frac{x_c r_p}{1+x_c r_p+y_c s_p+z_c t_p},\\
p_{cp}^\leftarrow&=&\frac{y_c s_p}{1+x_c r_p+y_c s_p+z_c t_p},\\
p_{cp}^\leftrightarrow&=&\frac{z_c t_p}{1+x_c r_p+y_c s_p+z_c t_p},\\
p_{cp}^\nleftrightarrow&=&\frac{1}{1+x_c r_p+y_c s_p+z_c t_p},
\end{eqnarray}
whose numerical value is determined by the following sufficient statistics, i.e. the reciprocal and non-reciprocal degrees of both core nodes

\begin{equation}
k_c^{\rightarrow}=\sum_p a_{cp}^{\rightarrow},\:k_c^{\leftarrow}=\sum_p a_{cp}^{\leftarrow},\:k_c^{\leftrightarrow}=\sum_p a_{cp}^{\leftrightarrow}
\end{equation}
(with $c=1\dots C$) and periphery nodes

\begin{equation}
h_p^{\rightarrow}=\sum_c a_{cp}^{\leftarrow},\:h_p^{\leftarrow}=\sum_c a_{cp}^{\rightarrow},\:h_p^{\leftrightarrow}=\sum_c a_{cp}^{\leftrightarrow}
\end{equation}
(with $p=1\dots P$). Notice that the binary variables defining $h_p^{\leftarrow}$ ($h_p^{\rightarrow}$) are the ones defining also $k_c^{\rightarrow}$ ($k_c^{\leftarrow}$): in fact, the non-reciprocated links outgoing from the core (periphery) are the same links incoming into the periphery (core). Finally, the estimation step for such a model reads

\begin{equation}
\left\{ \begin{array}{ll}
k_c^{\rightarrow}&=\:\:\:\sum_pp_{cp}^\rightarrow,\:\forall\:c,\\
h_p^{\rightarrow}&=\:\:\:\sum_cp_{cp}^\leftarrow,\:\forall\:p,\\
k_c^{\leftarrow}&=\:\:\:\sum_pp_{cp}^\leftarrow,\:\forall\:c,\\
h_p^{\leftarrow}&=\:\:\:\sum_cp_{cp}^\rightarrow,\:\forall\:p,\\
k_c^{\leftrightarrow}&=\:\:\:\sum_pp_{cp}^\leftrightarrow,\:\forall\:c,\\
h_p^{\leftrightarrow}&=\:\:\:\sum_cp_{cp}^\leftrightarrow,\:\forall\:p.
\end{array}
\right.\\
\label{sys2}
\end{equation}

\section*{Appendix C}

The aim of this appendix is providing simple examples of network configurations to further illustrate the methodology presented in the paper.

To this aim let us consider a bimodular structure where the link density of the two communities (whose number of nodes is $N_1=10$ and $N_2=90$ respectively) amounts at $p=q=0.8$ and where the two off-diagonal blocks have the same link density, i.e. $p^\top=p^\bot=r$. Let us, now, compare the explanatory power of the SBM and the DRG. The explicit calculation of the BIC for the SBM leads to the expression

\begin{eqnarray}
\text{BIC}_{SBM}=&-&2[N_1(N_1-1)(p\ln p+(1-p)\ln (1-p))+\nonumber\\
&+&N_2(N_2-1)(q\ln q+(1-q)\ln (1-q))+\nonumber\\
&+&2N_1N_2(r\ln r+(1-r)\ln (1-r))]+\nonumber\\
&+&4\ln [N(N-1)];
\end{eqnarray}
for consistency, the BIC for the DRG reads

\begin{eqnarray}\label{bic1}
\text{BIC}_{DRG}=&-&2N(N-1)[\overline{p}\ln\overline{p}+(1-\overline{p})\ln(1-\overline{p})]+\nonumber\\
&+&\ln [N(N-1)]
\end{eqnarray}
with $\overline{p}$ being the weighted average of the SBM probability coefficients. In fact,

\begin{eqnarray}
\langle L\rangle_{SBM}&=&N_1(N_1-1)p+N_2(N_2-1)q+2N_1N_2r\equiv\nonumber\\
&\equiv&\langle L\rangle_{DRG}=N(N-1)\overline{p}.
\end{eqnarray}

Let us now plot the trends of $\text{BIC}_{SBM}$ and $\text{BIC}_{DRG}$ as the parameter $r$ varies. As fig. \ref{fig6} shows, a region of values around $r=0.8$ exists where the SBM (i.e. the model specifying the network partition into modules) is penalized: notice, in fact, that the first terms of the two expressions coincide but the SBM correction term is larger than the DRG correction term. In other words, the network is homogeneous enough to be satisfactorily described by the only, global, parameter defining the DRG.\\

Let us now consider a core-periphery structure where the link density of the two communities (whose number of nodes is $N_1=10$ and $N_2=90$ respectively) amounts at $p^\bullet=0.8$ and $p^\circ=0.3$ and where the two off-diagonal blocks have the same link density, i.e. $p^\top=p^\bot=r$. Analogously to the previous example,

\begin{eqnarray}
\text{BIC}_{SBM}=&-&2[N_1(N_1-1)(p^\bullet\ln p^\bullet+(1-p^\bullet)\ln (1-p^\bullet))+\nonumber\\
&+&N_2(N_2-1)(p^\circ\ln p^\circ+(1-p^\circ)\ln (1-p^\circ))+\nonumber\\
&+&2N_1N_2(r\ln r+(1-r)\ln (1-r))]+\nonumber\\
&+&4\ln [N(N-1)];
\end{eqnarray}
while the BIC for the DRG is formally analogous to eq. \ref{bic1}. Plotting the trends of $\text{BIC}_{SBM}$ and $\text{BIC}_{DRG}$ as the parameter $r$ varies reveals that the SBM is always to be preferred. In this case, in fact, the network heterogeneity can never be accounted for by a single, global, parameter.\\

\begin{figure}[t!]
\begin{center}
\includegraphics[width=0.49\textwidth]{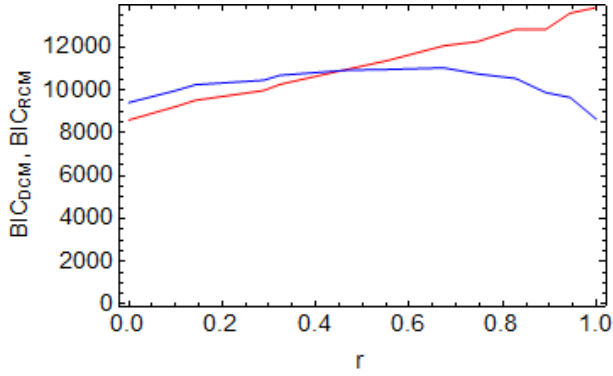}
\caption{Comparison between the numerical values of the BIC computed for the RCM (blue trend) and the DCM (red trend) on a network with an increasing level of reciprocity $r$.\label{fig7}}
\end{center}
\end{figure}

As a last case-study, let us now consider the comparison between the DCM and the RCM. To this aim, let us explicitly solve both models on binary, directed networks with an increasing level of reciprocity $r=\frac{\sum_{i\neq j}a_{ij}a_{ji}}{L}$. As fig. \ref{fig7} shows, as $r$ rises the performance of the RCM becomes increasingly preferable. To better understand this result, let us think about the two extreme configurations, i.e. the perfectly a-reciprocal one - with $r=0$ - and the perfectly reciprocal one - with $r=1$. In the first case, the evidence that $k_i^{out}=k_i^\rightarrow,\:\forall\:i$, $k_i^{in}=k_i^\leftarrow,\:\forall\:i$, $k_i^\leftrightarrow=0,\:\forall\:i$ induces probability coefficients satisfying the equalities $p_{ij}\simeq p_{ij}^{\rightarrow}$ and $p_{ji}\simeq p_{ij}^{\leftarrow},\:\forall\:i\neq j$, thus leading the DCM to be preferred. In the second case, the evidence that $k_i^{out}=k_i^{in}=k_i^\leftrightarrow\:\forall\:i$ induces probability coefficients satisfying the equalities $p_{ij}=p_{ji}=p_{ij}^{\leftrightarrow},\:\forall\:i\neq j$, thus leading the RCM to be preferred.

\section*{Acknowledgements}

This work was supported by the EU projects CoeGSS (grant num. 676547), DOLFINS (grant num. 640772), MULTIPLEX (grant num. 317532), Openmaker (grant num. 687941), SoBigData (grant num. 654024). RDC, as Newton International Fellow of the Royal Society, acknowledges support from the Royal Society, the British Academy and the Academy of Medical Sciences (Newton International Fellowship, NF170505).

\section*{Authors Contributions}

JLJ, RDC, GC, FS and TS developed the method. JLJ performed the analysis. JLJ, RDC, GC, FS and TS wrote the manuscript. All authors reviewed and approved the manuscript.

\section*{Additional Information}

The authors declare no competing financial interests.

\end{document}